# PORTLET WRAPPERS USING JAVASCRIPT

**ABSTRACT**

In this paper we extend the classical portal (with static portlets) design with HTML DOM Web clipping on the client browser using dynamic JavaScript portlets: the portal server supplies the user/passwords for all services through https and the client browser retrieves web pages and cuts/selects/changes the desired parts using paths (XPath) in the Web page structure. This operation brings along a set of advantages: dynamic wrapping of existing legacy websites in the client browser, the reloading of only changed portlets instead of whole portal, low bandwidth on the server, the elimination of re-writing the URL links in the portal, and last but not least, a support for Java applets in portlets by putting the login cookies on the client browser. Our solution is compliant with JSR168 Portlet Specification allowing portability across all vendor platforms.

## 1. INTRODUCTION

A portal provides a solution for aggregating content and applications from various systems into a single unified Web presentation to the end user and for conducting rich conversations in parallel. A few examples of well known portals include: Google Homepage, My Yahoo, MSN Live, My Jeeves.

One difficult and time-consuming job of the portal developers is to add interfaces for all of the systems that provide the external services into *portlets* or by using a service-oriented architecture with exposed web services: Web Services for Remote Portlets (WSRP). A *portlets* is an individual component displayed in the portal following the JSR168 Portlet Specification to allow portability across all vendor platforms. Practically, using these technologies, the software developer does not always separate cleanly into presentation and business logic layers.

In the following sections, we present our novel method of integrating existing Web pages with a single portal application by using dynamic client-side Web page scrapping portlets. We achieved this task by using a JavaScript (AJAX) [Garrett, J.J., 2005, Sun Microsystems, 2005] portlet to retrieve the source Web pages and to Web clip the desired parts on the client browser using Javascript and HTML DOM. The interaction with the portal requires re-loading only of changed portlets instead of whole portal, determining a gain in Web page retrieval. As a result, we developed a portal application with an easy to use tool, which permits to the user can create portlets which clip from existing Web pages desired parts into portlets. A series experiments have been conducted to evaluate the performance of our portal solution.

## 2. METHODS, RESULTS AND RELATED WORK

Our design method integrates wrapping portlets in portals at the convenience of providing the URL, security credentials and the wrapping method ()XPath. The portal server supplies the location (URL), the security credentials (the HTML fields and the user/passwords) and the wrapping elements (using XPATH) (previously provided by the user) for all services through HTTPS whereas the client browser retrieves the web pages and cuts/selects/changes the desired parts and wraps them into a portlet on the portal page (see Figure 1a for the portal initialization). After the initial contact with the server, the client can communicate only with the producing websites individually, based on the interaction with the user and a predefined portal workflow without portal server interaction (see Figure1b).

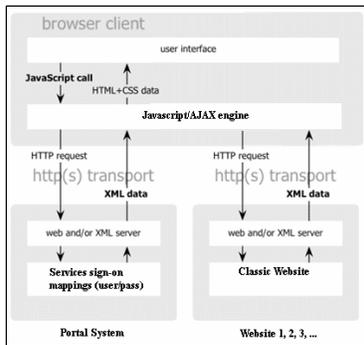 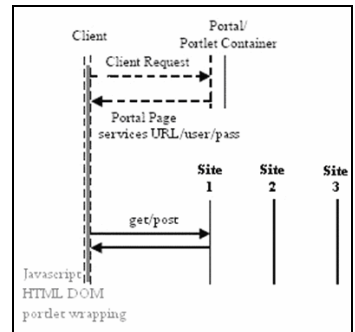

a. Portal with portlet scrapping using client-side JavaScript;  b. Client-side interaction progress

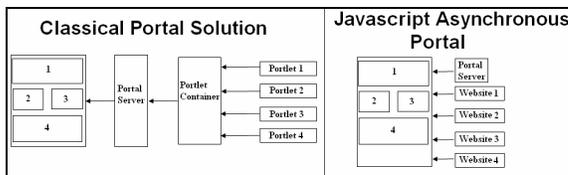 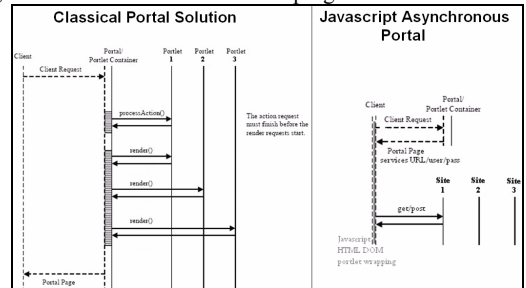

c. Comparison components diagram of portal solutions;   d. Comparison of interaction diagram for portals

Figure 1. Dynamic Javascript Asynchronous portal: components, interaction diagram, and comparisons

A comparison of our design with the classical portal solution (see Figure 2c and 2d) show that by using our design pattern we solve some of the portal problems: we achieve content aggregation for legacy applications by dynamically wrapping existing websites in the client browser and the personalization is defined in the portal structure and inherited from each service websites, employ low bandwidth on the server, and reload only the changed portlets instead of whole portal. Our solution eliminates re-writing the URL links in the portal, supports Java applets in the portlets by putting the login cookies on the client and not on the server, and is compliant with the JSR168 Portlet specification (which allows portability to other portal platforms).

Based on this architecture, we implemented a tool through which the user supplies the location (URL), the security credentials (the HTML fields and the user/passwords) and the wrapping elements (using XPATH) for each portlet in the portal. The portal works following a previously defined protocol to retrieve web pages, clips the desired parts and wraps them in a portlet on the portal page (see example from Figure 2 for a college faculty/student academic portal).

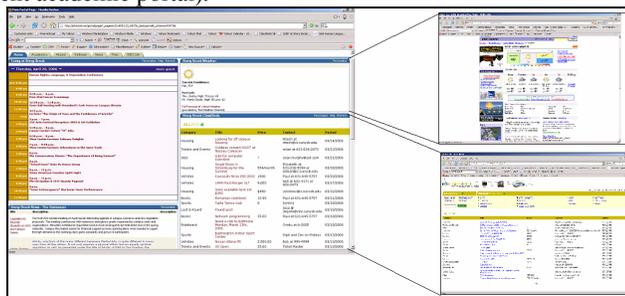

Figure 2. A faculty/student academic portal resulted from client side Web clipping

Parts of this work have been done before, but they have never been integrated into a single solution. Oracle Application Server Portal 10g [Oracle Application Server Portal 10g User Manual, 2005] does server-side clipping of existing Web pages, but the server follows the component diagram for classical portal solutions from Figure 4 which puts the portal container on the portal server between the client and the clipped source Web sites (and by this losing some of our advantages). Asynchronous Rendering of Portlet Content with AJAX [Ziebold, G., Suri, J., Sum M., 2005] and AJAX tags [Mitchell J., Husted, T., Brown, D., 2005] deal only with the problem of loading individual portlets, but it doesn't do automatic scraping.

## 3. CONCLUSION

In this paper, we have described the design of client-side wrappers for portlets, a client-side dynamic portlet solution. We motivated how this design makes the portal easily adaptable, improve flexibility, offer functionality for legacy websites and are compliant with existent wide-used standards. The following are only a few potentially useful areas for further research. Currently, the desired paths (XPath) in the source Web sites are manually coded. If the structure of the source Web sites is semantically annotated and the relevant paths are identified, our algorithm can be easily extended to allow smart path detection within a Web page. NLP techniques can be employed to enhance substructure processing and searching. We are also currently researching the feasibility of applying machine learning algorithms and statistical models to identification and ranking of desired paths in the structure of the sources. Our portal system can be further enhanced by making portlets connected with the overall portal business workflow.